\documentclass{ws-procs975x65}

\begin{document}

\markboth{Gergely, Keresztes, Mik\'oczi}
{The second post-Newtonian order generalized Kepler equation}

\title{THE SECOND POST-NEWTONIAN ORDER GENERALIZED KEPLER EQUATION
\footnote{
Research supported by OTKA grants no. T046939, TS044665 and the J\'{a}nos
Bolyai Fellowships of the Hungarian Academy of Sciences. L.\'A.G. wishes to
thank the organizers of the 11th Marcel Grossmann Meeting for support.} 
}

\author{L\'ASZL\'O \'A. GERGELY, ZOLT\'AN KERESZTES and BAL\'AZS MIK\'OCZI}

\address{Departments of Theoretical and Experimental Physics, 
University of Szeged,\\
D\'om t\'er 9, H-6720 Szeged, Hungary\\
\email{gergely@physx.u-szeged.hu, zkeresztes@titan.physx.u-szeged.hu, 
mikoczi@titan.physx.u-szeged.hu}}

\begin{abstract}
The radial component of the motion of compact binary systems 
composed of neutron stars and/or black holes on eccentric orbit 
is integrated. We consider all type of perturbations that emerge 
up to second post-Newtonian order. These perturbations are either 
of relativistic origin or are related to the spin, mass quadrupole 
and magnetic dipole moments of the binary components. We derive a 
generalized Kepler equation and investigate its domain of validity,
in which it properly describes the radial motion. 
\end{abstract}


\bodymatter

\section*{}

Compact binaries composed of neutron stars / black holes are radiating
gravitational waves. The waveform and phase of gravitational waves are
strongly influenced by the orbital evolution of these systems. Before the
system reaches the innermost stable orbit, its evolution can be well
described by a post-Newtonian (PN) expansion about the Kepler motion. As
dissipative effects due to gravitational radiation only enter at 2.5 PN
orders, the orbital evolution is conservative up to the 2PN\ orders. Even to
this order the dynamics is complicated enough not only by the general
relativistic corrections to be added at both the first and second PN orders
and by tail effects, but by \textit{finite size effects} as well. These
include spins, mass quadrupolar and magnetic dipolar moments.

From among these the spin is the dominant characteristic. The effect of the
spin-orbit coupling on the motion has been considered long time ago,
\cite{BOC} and revisited more recently.\cite{KWW,Kidder,RS,GPV1,GPV2,GPV3,FBB,BBF} 
This contribution suffers from the non-uniqueness in the definition of the 
spins, expressed by the existence of
at least three different spin supplementary conditions. The physical results
however should be independent of the chosen SSC.

The next contributions (at 2PN) are due to spin-spin coupling. 
\cite{BOC,Kidder,spinspin1,spinspin2,self} These include proper spin-spin
contributions between the two components as well as spin self-interactions.
An effect of similar size is due to the mass quadrupoles of the binary
components. This is the so-called quadrupole-monopole 
interaction,\cite{BOC,Poisson,quadrup} representing the effect on 
the motion of one of the
components (seen as a test mass) in the quadrupolar field of the other
component. The quadrupole moment may either be a consequence of rotation or
it may be not. As magnetars with considerable magnetic field are known, the
possibility of the coupling between the magnetic dipole moments was also
investigated.\cite{IT,mdipole} With both components having the magnetic
field of $10^{16}$ Gauss, the magnetic dipolar contribution provides other
2PN contributions to the dynamics.

Although the above enlisted effects emerge either at 1.5 PN (spin-orbit) or
at 2PN orders (spin-spin, quadrupole, possibly magnetic dipole), they all
represent the leading order contributions of the respective type. In this
sense they are linear perturbations of the Keplerian motion. For these
perturbations the radial part in the motion of a compact binary system
decouples from the angular motion. With the aid of the turning points of the
radial motion, given as $\dot{r}=0$ both a radial period and suitably
generalized true and eccentric anomaly parametrizations of the radial motion
can be derived \cite{param,Gparam}. The eccentric anomaly parameter $\xi $
agrees with the corresponding parameter $u$ of the Damour-Deruelle formalism 
\cite{DD}. The true anomaly parameter $\chi $ however is different from the
parameter $v$. The complex counterparts of these parametrizations have the
wonderful property that the overwhelming majority of the radial integrals
can be evaluated simply as the residues in the origin of the complex
parameter plane.

Employing these convenient parametrizations $u$ and $\chi $, the radial
motion could be integrated exactly. The result is a generalized Kepler
equation\cite{Kepler}:
\begin{eqnarray}
n\left( t-t_{0}\right) &=&\xi -e_{t}\sin \xi +F\left( \chi ;\psi _{0},\psi
_{i}\right) ~,  \nonumber \\
F\left( \chi ;\psi _{0},\psi _{i}\right) &=&f_{t}\sin \left[ \chi +2\left(
\psi _{0}-\overline{\psi }\right) \right] +\sum_{i=1}^{2}f_{t}^{i}\sin \left[
\chi +2\left( \psi _{0}-\psi _{i}\right) \right] \ ,  \label{Kepler}
\end{eqnarray}%
where $n$, $e_{t}$, $f_{t}$ and $f_{t}^{i}$ are orbital elements. 
Most notably, the true anomaly parametrization 
$\chi $ appears only in combination with the coefficients $f_{t}$ and $%
f_{t}^{i}$, which in turn receive contributions only from spin-spin, mass
quadrupolar and magnetic dipolar contributions. These terms also contain the
azimuthal angles $\psi _{i}$ of the spins (with $2\overline{\psi }=\psi
_{1}+\psi _{2}$). The angle $\psi _{0}$ is the argument of the periastron
(the angle subtended by the periastron and the intersection line of the
planes perpendicular to the total and orbital angular momenta, respectively).

Besides the convenient parametrization and integration relying on the use of
the residue theorem, the other main ingredient in obtaining the result (\ref%
{Kepler}) was the introduction of averaged dynamic quantities $\overline{A}
$ and $\overline{L}$. These represent averages of the magnitudes of the
Laplace-Runge-Lenz and orbital angular momentum vectors, respectively. The
averages are taken over the angular range defined by one radial period.
Although the quantities $A$ and $L$ are not constant under the spin-spin,
quadrupole and magnetic dipole couplings, their angular average over a
radial period remarkably is (as long as we are considering conservative
dynamics).

Another important point to stress is that the orbital elements from Eq. 
(\ref{Kepler}) depend on the relative angle $\gamma$ between the spins and the 
angles $\kappa_{1,2}$ of the spins span with the orbital angular momentum. 
These in turn evolve, bearing a hidden time-dependence. 
However, the precessional motion due to the spin-orbit coupling does not affect 
them, while the error made by disregarding the changes due to 
the spin-spin interaction are quite small. To see this we note that the lowest
order in which $\kappa_{1,2}$ occur in Eq. (\ref{Kepler}) are the spin-orbit terms 
at 1.5 PN. 
Their change being an 1.5 PN effect, \cite{GPV3} a variation  
appears only at 3PN accuracy in the Kepler equation. 
The change in $\gamma$ is at 1PN, \cite{GPV3} however $\gamma$ 
enters only in the spin-spin contributions, its change becoming
significant therefore again at 3PN accuracy. These are
smaller effects (appearing at 0.5 PN higher order) than those occurring from the 
leading order radiation reaction. 
Nevertheless, such changes accumulate over the inspiral.
Therefore the Kepler equation (\ref{Kepler}) with \textit{constant}
coefficients should be applied with care. As the magnitude of the still
disregarded effects depends on the value of the post-Newtonian parameter $%
\varepsilon =Gm/c^{2}r=v^{2}/c^{2}$, they are higher during the last stages
of the inspiral. In conclusion the Kepler equation with constant
coefficients, Eq. (\ref{Kepler}) represents a better approximation in the
early stages of the inspiral.

In order to include the general relativistic 2PN contributions, the 2PN
terms\cite{DS87,DS88,SW93,NW95} given in terms of the $v$-parametrization 
should be also added to the Kepler equation. Such a Kepler equation
represents the complete solution of the radial motion to 2PN orders. We
conclude with the remark that a full parametrization of the radial motion up
to 2PN orders, with the inclusion of finite-size effects is possible with
the ensemble of three radial parameters $\left( u\equiv \xi ,~v,~\chi
\right) $.

\bibliographystyle{ws-procs975x65}

\end{document}